\documentclass[twocolumn,tighten]{aastex631}

\usepackage{graphicx}
\usepackage{appendix}
\usepackage{hyperref}

\received{July 10, 2022}
\revised{September 19, 2022}
\accepted{September 20, 2022}

\shorttitle{Echoes of light revealed by HST}
\shortauthors{Stritzinger, Taddia et al.}

\begin{document}
\title{\textit{Hubble Space Telescope} Reveals Spectacular Light Echoes Associated with the Stripped-envelope Supernova~2016adj in the Iconic Dust Lane of Centaurus~A}

\correspondingauthor{M.~D. Stritzinger}
\email{max@phys.au.dk}

\author[0000-0002-5571-1833]{Maximilian D. Stritzinger}
\affil{Department of Physics and Astronomy, Aarhus University, Ny Munkegade 120, DK-8000 Aarhus C, Denmark}

\author[0000-0002-2387-6801]{Francesco Taddia}
\affiliation{Department of Physics and Astronomy, Aarhus University, Ny Munkegade 120, DK-8000 Aarhus C, Denmark}

\author[0000-0002-7491-7052]{Stephen S. Lawrence}
\affiliation{Dept. of Physics \& Astronomy, 151 Hofstra University, Hempstead, NY 11549, USA}

\author[0000-0002-0537-3573]{F. Patat}
\affiliation{European Organization for Astronomical Research in the Southern Hemisphere (ESO), Karl-Schwarzschild-Str. 2, 85748, Garching b. München, Germany}

\author[0000-0003-2191-1674]{Morgan Fraser}
\affiliation{School of Physics, O'Brien Centre for Science North, University College Dublin, Belfield, Dublin 4, Ireland}

\author[0000-0002-1296-6887]{Llu\'{i}s Galbany}
\affiliation{Institute of Space Sciences (ICE, CSIC), Campus UAB, Carrer de Can Magrans, s/n, E-08193 Barcelona, Spain}
\affiliation{Institut d’Estudis Espacials de Catalunya (IEEC), E-08034 Barcelona, Spain}

\author[0000-0002-3415-322X]{Simon Holmbo}
\affiliation{Department of Physics and Astronomy, Aarhus University, Ny Munkegade 120, DK-8000 Aarhus C, Denmark}

\author[0000-0002-3350-4243]{Ali Hyder}
\affiliation{Department of Astronomy, New Mexico State University, PO BOX 30001, MSC 4500, Las Cruces, NM 88003-8001, USA}
\affiliation{Dept. of Physics \& Astronomy, 151 Hofstra University, Hempstead, NY 11549, USA}

\author[0000-0001-6209-838X]{Emir Karamehmetoglu}
\affiliation{Department of Physics and Astronomy, Aarhus University, Ny Munkegade 120, DK-8000 Aarhus C, Denmark}

\begin{abstract}
We present a multi-band sequence of \textit{Hubble Space Telescope} images documenting  the emergence and evolution of multiple light echoes (LEs) linked to the stripped-envelope supernova (SN) 2016adj located in the central dust-lane of Centaurus~A. Following point-spread function subtraction, we identify the earliest LE emission associated with a SN at only $+$34 days (d) past the epoch of $B$-band maximum. Additional HST images extending through $+$578~d cover the evolution of LE1 taking the form of a ring, while images taken on $+$1991~d reveals not only LE1, but also segments of a new inner LE ring (LE2) as well as two additional outer LE rings (LE3 \& LE4). Adopting the single scattering formalism, the angular radii of the LEs suggest they originate from discrete dust sheets in the foreground of the SN. This information, combined with measurements of color and optical depth of the scattering surfaces, informs a scenario with multiple sheets of clumpy dust characterized by a varying degree of holes. In this case, the larger the LE's angular radii, the further in the foreground of the SN its dust sheet is located. However, an exception to this is LE2, which is formed by a dust sheet located in closer proximity to the SN than the dust sheets producing LE1, LE3, and LE4. The delayed appearance of LE2 can be attributed to its dust sheet having a significant hole along the line-of-sight between the SN and Earth.

\end{abstract}


\section{Introduction} 
\label{sec:intro}

Light echoes (LE) are produced when photons emitted by a transient source scatter off interstellar dust clouds \citep[see][and references therein]{Sugerman2003}.
Often appearing as arcs or wisps of light, LEs were first detected in Galactic and extra-galactic novae over a hundred years ago \citep{kapteyn1901,ritchey1901,swope1940}, and successfully explained as scattering phenomena by \citet{couderc1939}. Later, LEs were found around variable stars \citep{havlen1972,bond2003,rest2012b}, while more recently, LEs associated with historical supernovae (SNe) have been documented \citep{rest2008,rest2011,vogt2012}.
Beginning with SN~1987A \citep{crotts1988,suntzeff1988}, LE emission has been documented in a dozen nearby SNe \citep[see][and references therein]{yang2017}. 

\begin{figure*}[!ht]
\begin{center}
\includegraphics[width=9.5cm,angle=270]{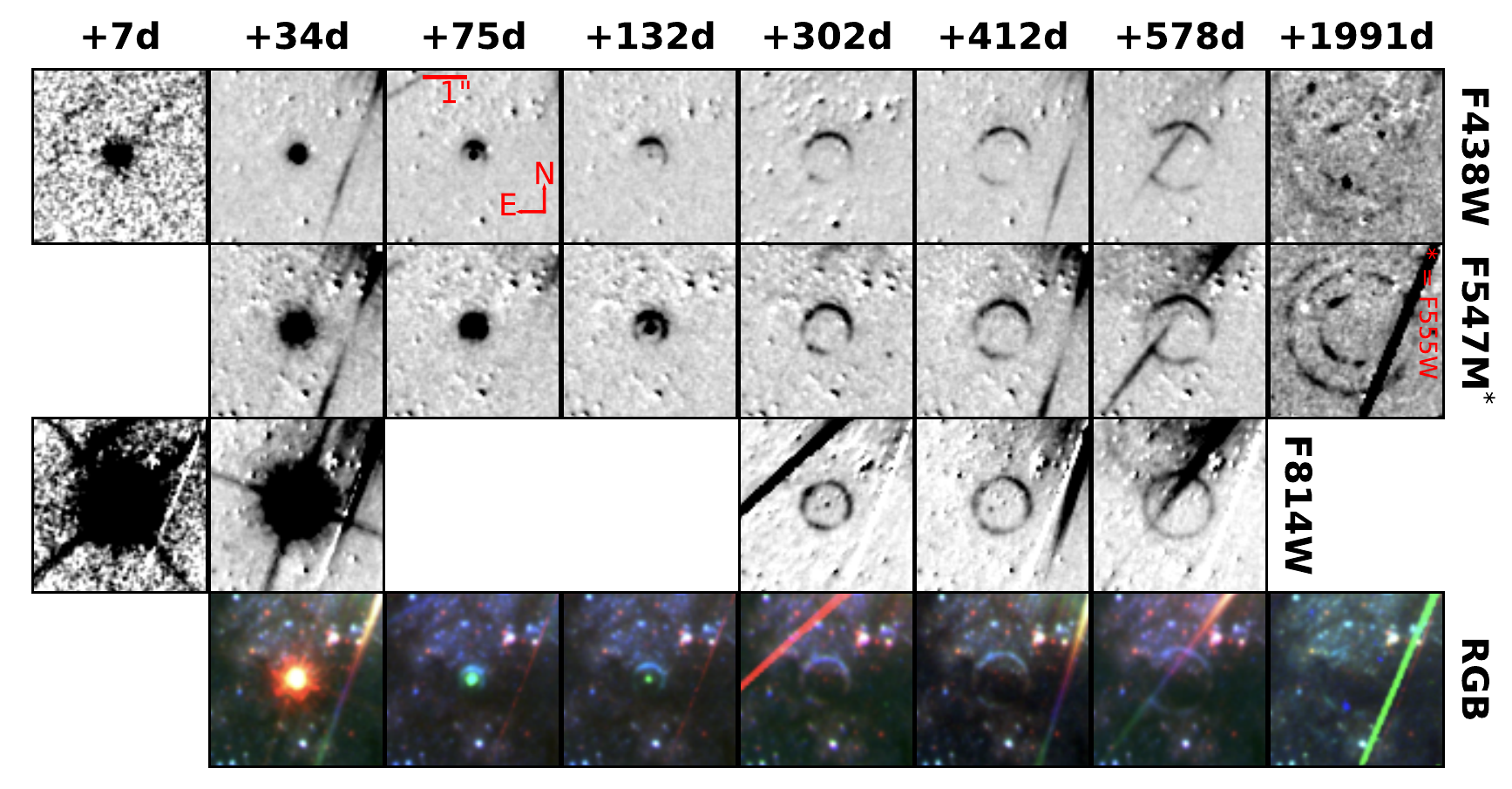}
\caption{A montage of galaxy-subtracted HST images of SN~2016adj extending $+$7~d to $+$1991~d post the epoch of $B$-band maximum. The $F438W$- \textit{(first row)}, $F547M$- \textit{(second row)}, and $F814W$-band \textit{(third row)} images reveal the emergence and evolution of multiple light echoes. Corresponding RGB composite images are presented in the bottom row. The LE (hereafter LE1) first appearing as a half-ring  in the $+$34~d data (see Appendix) becomes visually apparent on $+75$~d, and continues to radially expand through $+$1991~d.  The most recent HST images also reveals a new partial inner LE ring (hereafter LE2), as well as LE segments of two additional outer rings located beyond LE1 and designated LE3 and LE4 (see also Fig.~\ref{fig:lateLE}). The SN position is centered in each of the ($4\farcs36 \times 4\farcs36$) panels and is no longer visually apparent by $+$302~d. Many of the images are contaminated with a conspicuous saturation spike associated with a foreground star. Finally, note that the SN is saturated in the first two $F814W$ images. }
\label{fig:echo}
\end{center}
\end{figure*}


In principle, measurements of the angular radius ($\alpha$) of a LE as projected on the sky, along with the surface brightness, color evolution, and polarization signatures, provides a means to infer the structure, geometry, and extinction properties of the underlying dust including grain size and composition \citep{couderc1939,chevalier1986,patat2005}. Studies of SN~1987A \citep{xu1995} and SN~2014J \citep{yang2017} reveal multi-component LEs consisting of bright arcs with knot-like structures superposed on more diffuse, full-ring emission. Reconstructed 3-D dust mappings of both objects suggests the arcs and rings are produced by discrete slabs of interstellar dust, characterized by different thicknesses, column densities, and extinction properties \citep{xu1995,yang2017}. 

We present a time-series of multi-band images taken with the \textit{Hubble Space Telescope} (HST) of SN~2016adj nestled in the iconic dust lane of NGC~5128; hereafter Cen~A \citep{Hyder2018}. The data documents four distinct LE components that emerge at different times and which are associated with distinct sheets of dust. Our analysis of the LEs characterize the locations and properties of the various sheets of dust producing the echoes. 

BOSS (Backyard Observatory Supernova Search) discovered SN~2016adj on 2016 February 08.56 UT (i.e., JD--2,457,427.06) with $m_V = 14$ mag \citep{marples2016,kiyota2016}. Spectra obtained soon after discovery indicated it was a core-collapse SN, with  reports differing in the sub-type, i.e., SN~II \citep{yi2016}, SN~Ib \citep{stritzinger2016}, or  SN~IIb \citep{hounsell2016,thomas2016}. \citet{banerjee2018} presented a near-infrared (NIR) spectral sequence of SN~2016adj and focused on modeling early carbon-monoxide emission. Although \citeauthor{banerjee2018} adopted a SN~IIb classification, they did note a lack of hydrogen and helium lines in their spectra. In a forthcoming paper (Stritzinger et al., in prep), we establish the optical/NIR spectra of SN~2016adj are fully consistent with those of carbon-rich SNe~Ic \citep{valenti2008,Shahbandeh2021}.

\section{HST observations of SN~2016adj}

HST observed SN~2016adj over eight epochs extending from 22 February 2016  to 28 July 2021 with the Wide Field Camera 3 (+WFC3). Drizzled HST frames were retrieved from the  Mikulski Archive for Space Telescopes,  consisting of 8, 6, and 5 epochs taken with the $F438W$, $F547M$, and $F814W$ filters, respectively, and a single epoch with the $F555W$ filter. A summary of the HST images used in this study is provided in Table~\ref{tab:hst_log}. This includes image IDs, program identification numbers, date of observations, phase relative to the epoch of $B$-band maximum, exposure times, and filter identifications. The dates of observations correspond to $+7$~days (d) and $+$1991~d past the epoch of $B$-band maximum.\footnote{The epoch of $B$-band maximum is estimated from a $B$-band light curve of SN~2016adj obtained by the \textit{Carnegie Supernova Project} to have occurred  on 15 February 2016 (i.e., JD--$2,457,433.5\pm2$; Stritzinger, M. et al., in prep).}
In addition to the HST science images, multi-band HST images of the host taken during 2010 were used to perform galaxy-template subtraction on each of the post-SN images. 

\begin{figure}[htb]
\centering
\includegraphics[width=0.95\linewidth,angle=270]{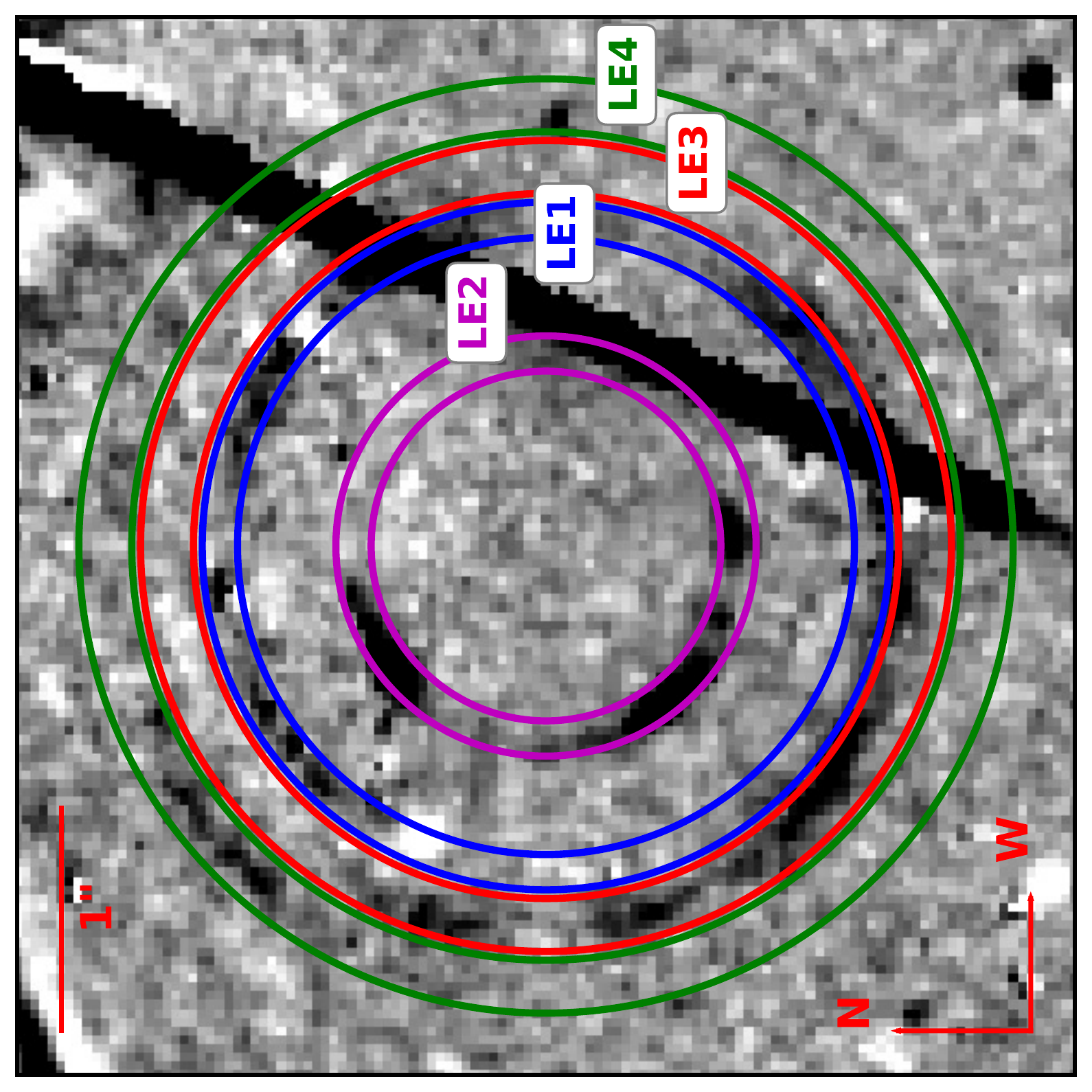}
\caption{Host-subtracted $F555W$-band HST image of SN~2016adj on $+$1991~d, with the positions of LE1, LE2, LE3, and LE4 highlighted with colored rings and labeled.}
\label{fig:lateLE}
\end{figure}
 
\begin{figure*}[!htb]
\centering
$\begin{array}{cc}
\includegraphics[width=8.8cm]{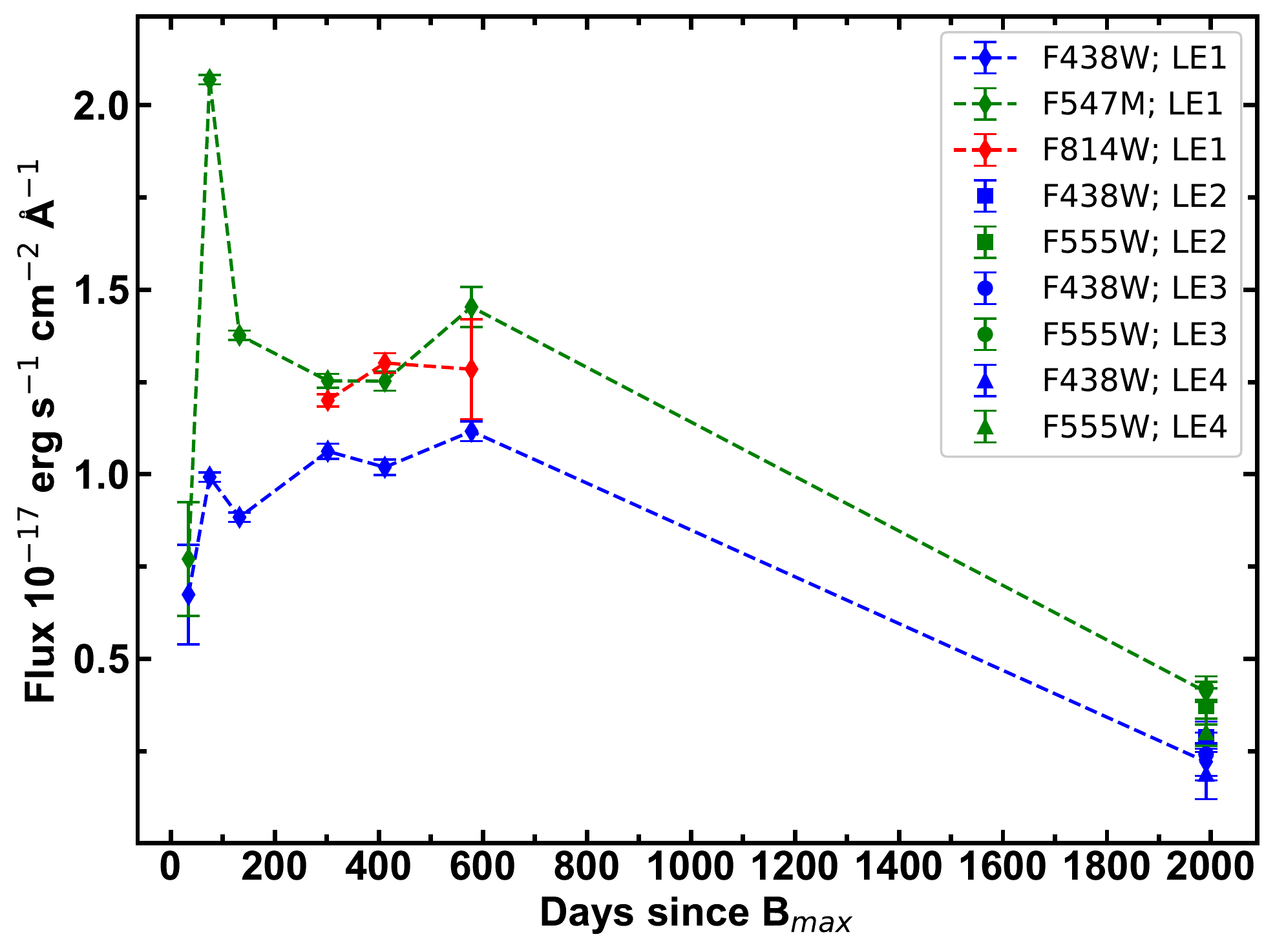}   & 
\includegraphics[width=8.8cm]{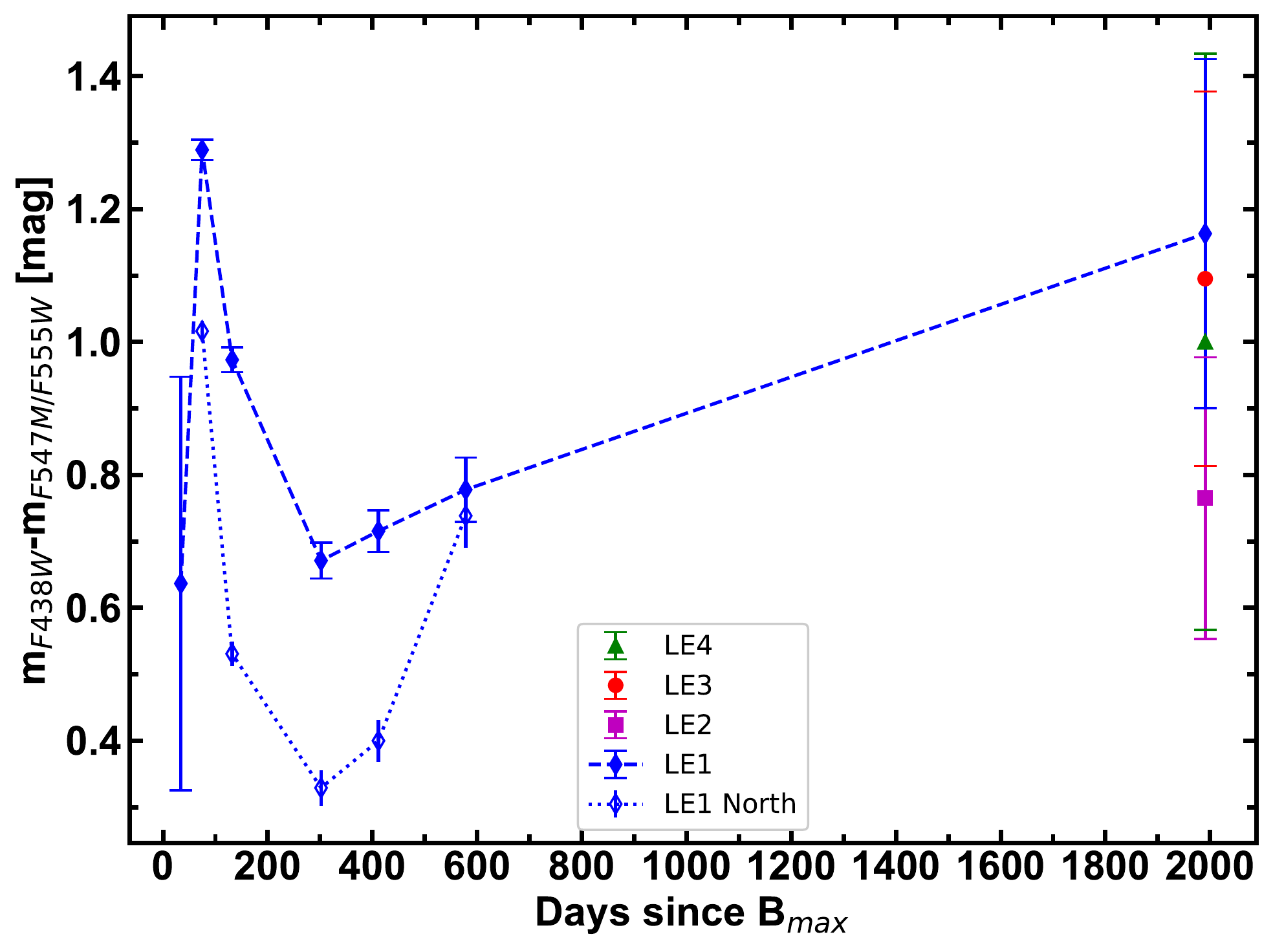}
\end{array}$
\caption{\textit{Left:} Multi-band integrated flux evolution of LE1, as well as flux measurements of LE2--LE4 on $+$1991~d. \textit{Right:} Apparent ($m_{F438W} - m_{F547M}$) color evolution of LE1, and colors of LE2-LE4 (last epoch corresponds to $m_{F438W} - m_{F555W}$). Also plotted is the color of LE1 estimated from the segment of its ring extending over the North direction and  which is clearly bluer (see Fig.~\ref{fig:LEdust}) than the color inferred from the total integrated flux.}
\label{fig:echo_flux}
\end{figure*}

\section{Results}
\label{sec:results}

\subsection{The light echoes of SN~2016adj}

A montage of template-subtracted HST images of SN~2016adj shown in Fig.~\ref{fig:echo} reveals four distinct LE components present at various epochs.\footnote{An animation of LE1 is available on  the electronic ApJ Letters version of this paper.} 
LE1 is first visually apparent as a half-ring on $+$75~d with a projected angular radius on the sky of $\alpha = 0\farcs27\pm0\farcs01$. The intensity of the half-ring LE is maximum at a position angle (PA) corresponding to the North, and decreases until almost disappearing towards the West and East directions (see Fig.~\ref{fig:fluxvsPA}). Following \citet{sugerman2016b}, LE1 is recovered in the $+$34~d $F438W$-band image as described in Appendix~\ref{sec:recoveringLE} and shown in Fig.~\ref{fig:psfsub}. To our knowledge, this is the earliest detection of a LE surrounding a SN. Turning to epochs of $+$132~d and later, while the SN has faded, LE1 continues to appear as a radially expanding ring with a non-uniform surface-brightness out to $+$1991~d. As highlighted in Fig.~\ref{fig:lateLE}, the final epoch reveals additional LE components including: an  `inner' LE ring labeled LE2 that lies inside LE1, and segments of two `outer' rings designated LE3 and LE4.

\begin{figure*}[!htb]
\centering
\includegraphics[width=12cm]{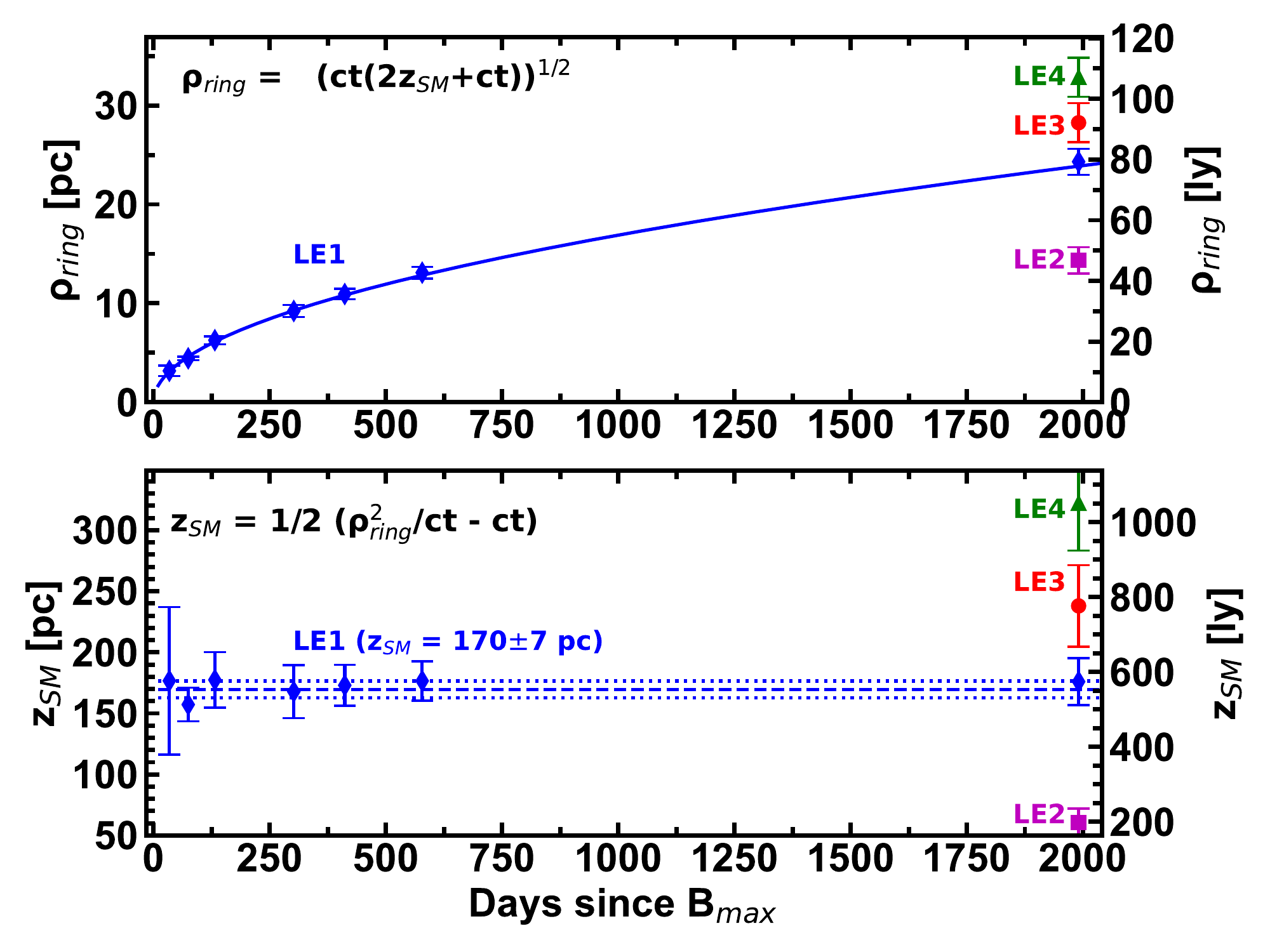} 
\caption{\textit{Top:} Values of $\rho_{ring}$ for LE1--LE4. The blue line corresponds to the relation provided in the panel assuming for LE1 $z_{SM} = 170$~pc. \textit{Bottom:} Values of $z_{SM}$ as inferred from Eq.~\ref{eq:zsm} for each LE component. The mean $z_{SM}$ value and the associated uncertainty are indicated with a dashed and dotted lines, respectively.}
\label{fig:echo_radius_time}
\end{figure*}

 Values of $\alpha$ for each LE component were measured following the procedure described in Appendix~\ref{sec:measuringangle}, and the results are listed in Table~\ref{tab:radiuslog}. 
A linear fit to $\alpha$ as measured for LE1 over seven epochs reveals an expansion velocity of 0\farcs018 per month. This translates to the superluminal value of $\sim 12c$, which is 0.75 times the value inferred for SN~1987A \citep{suntzeff1988}. Furthermore, by fitting $\alpha$ between the first three epochs, we obtain an impressive superluminal value of $\sim 37c$.

\subsection{Integrated brightness and color evolution}

To measure the LE brightness from the montage of data the point-spread function (PSF) of the SN was first removed from the $F438W$ and $F547M$ images taken on $+$34~d and $+$75~d. PSF subtraction was performed making use of tools within the \texttt{Astropy  photoutils} library \citep{bradley2020}. To do so, a model PSF was constructed based on the PSFs of two-dozen isolated field stars. The model PSF was scaled to the peak  SN and LE signal, and then subtracted, revealing the LE signal.  
Next, template images from 2010 were subtracted from all science images, and then photometry was computed for each LE component using two apertures centered on the SN position with radii respectively larger and smaller than the LE radius (e.g., Fig.~\ref{fig:psfsub}). The flux level of the innermost aperture was subtracted from that of the outermost aperture, yielding the LE flux. Integrated LE flux measurements (and magnitudes) of LE1--LE4 determined from the entire sequence of images is listed in Table~\ref{tab:radiuslog}, with uncertainties derived from the standard deviation of the background next to the ring. Flux values and the apparent colors for each LE component are plotted in Fig.~\ref{fig:echo_flux}. 

Inspection of the left panel of Fig.~\ref{fig:echo_flux} reveals that on $+$34~d LE1 exhibits similar flux values in both bands. Interestingly, while by $+75$~d the flux in the $F438W$ band only slightly increases, in the  $F547M$ band its brightness increases by $1.3\pm0.2$ mag. This results in the color evolving to the red over the same time period reaching a value of 1.3 mag on $+75$~d.  Beyond $+$75~d and  $+$302~d as the flux in the $F547M$ decreases and that of the $F438W$ marginally increases, the color of LE1 returns back to the blue reaching a value similar to that  measured on $+$34~d. Over the remaining duration of the observations and as the LE brightness decreases in both the $F438W$ and $F547M$, the color of LE1 slowly evolves back towards the red.  By $+$1991~d the brightness and colors of LE1 is similar to that measured for LE3 and LE4, although LE2 appears marginally bluer.

 \subsection{Foreground distance of the scattering dust sheets}

Following the single scattering formalism \citep{couderc1939,patat2005}, the distance ($z_{SM}$) between a transient light source and a foreground LE producing dust sheet can be inferred from the geometric relationship 

\begin{equation}
\label{eq:zsm}
z_{SM} = {1\over2}\left({{\rho_{ring}^2}\over{ct}} - ct\right) .
\end{equation}
\noindent Here $z_{SM}$ is the foreground distance to the scattering medium along the line-of-sight (LoS) in parsecs, $c$ is the speed of light, $t$ is the time since $B$-band maximum, and $\rho_{ring}$ corresponds to the apparent ring radius in parsecs as projected on the sky and determined from the $\alpha$ values assuming the Cepheid distance to Cen~A of $3.42\pm0.18$ (random) $\pm$0.25 (systematic) Mpc \citep{ferrarese07}. 
Measurements of $\rho_{ring}$ and the inferred $z_{SM}$ values for each LE component are plotted in Fig.~\ref{fig:echo_radius_time} and listed in Table~\ref{tab:radiuslog}. The $z_{SM}$ values suggests the LEs are produced by dust sheets contained within a foreground volume extending between $z_{SM} \sim 60-320$ pc. 

\citet[][their Eq.~11]{Sugerman2003} provides a relation to estimate the thickness of a scattering dust sheet ($\Delta z_{SM}$) in terms of $\rho$, $\Delta \rho$, $t$, and a SN light-curve width parameter $\tau$. Following \S~\ref{sec:deltaz}, we obtain for LE1 $\Delta z$ estimates between 68-117~pc, while for LE2, LE3 and LE4   $\Delta z$ values of $16\pm5$ pc, $32\pm11$ pc, and $38\pm13$ pc, respectively. These values along with the compactness of the  radial profiles of each LE,  suggests they are formed by a single scattering process that occurs in distinct dust sheets \citep{patat2005,yang2017}.

 \section{Discussion}
 \label{sec:discussion}

The color evolution of LE1 is perplexing, though not  unexpected given the totality of the dust lanes of Cen~A as highlighted in Fig.~\ref{fig:LEdust}.
We first turn to understand the color evolution in terms of the $V$-band effective optical depth ($\tau^V_{eff}$) of the scattering dust sheets, while later consequences of dust reddening on the color evolution is considered. In doing so, relatively high $\tau^V_{eff}$ values are computed, however, after accounting for reddening the values are consistent with the single scattering plus attenuation parameter space considered by \citet{patat2005}.

 \citet[][their Eq.~21]{patat2005} relates the observed LE color $(B-V)_{LE}$ to the intrinsic color $(B-V)_{SN_0,peak}$ of the SN at peak following: $\tau^V_{eff} \approx 2.8 \left[(B-V)_{LE} - (B-V)_{SN_0,peak}\right]$.
 Given the high and uncertain reddening parameters associated with SN~2016adj (Stritzinger et al. in prep), we adopt $(B-V)_{SN_0} \approx 0.3\pm0.05$ mag \citep[see][their Fig. 5]{stritzinger2018}, which implies $\tau^V_{eff}$ values of $1.0\pm1.0$ to $2.8\pm0.2$ (see Table~\ref{tab:radiuslog}). 
 
 \begin{figure}[!htb]
\includegraphics[width=8cm]{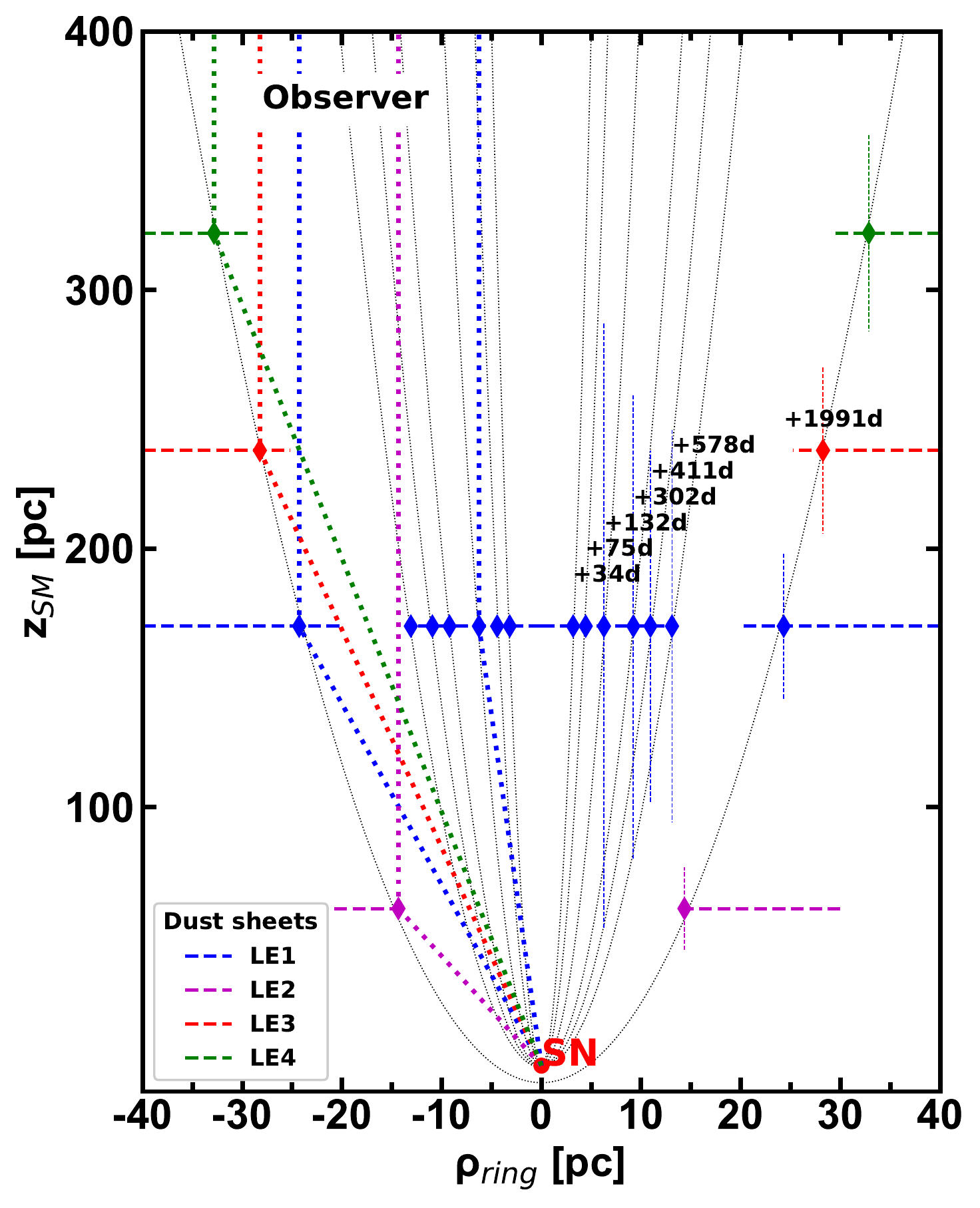}
\caption{Schematic of multiple sheets of clumpy dust located along the line-of-sight to Earth creating the various LE signatures of SN~2016adj. The parabolas indicate surfaces of equal light travel time to the observer. The horizontal dashed lines mark $z_{SM}$ inferred for the dust sheets producing the LEs, and vertical dotted lines indicate  their estimated widths ($\Delta z_{SM}$). Note that the sheets of dust do not  necessarily  extend symmetrically on both sides of the LoS.}
\label{fig:LEschematic}
\end{figure}

 Returning to the apparent ($B-V$) color evolution in Fig.~\ref{fig:echo_flux} of LE1 between epochs 1 and 2 and epochs 3 and 4, we find the abrupt changes in color is associated with changes in $\tau^V_{eff}$. For example, LE1 abruptly evolves to the red as $\tau^V_{eff}$ increases from $\sim 1.0\pm1.0$ to $2.8\pm0.2$. Similarly, the evolution of LE1 to the blue between epoch 2 and 4 occurs as $\tau^V_{eff}$ decreases from $\sim 2.8\pm0.2$ to $1.1\pm0.2$. The color evolution to the blue  could be due to changes of $\tau^V_{eff}$ and not due to geometrical effects coupled to the various wavelengths dependencies. Such evolution is expected with steep density gradients and/or fluctuations in the dust distributed along the LoS, which leads to less self-absorption of blue photons, and hence brighter/bluer LE emission \citep[see][]{patat2005}.

The previous results assume that the apparent LE color is not affected by host-galaxy reddening. However, as indicated by  Fig.~\ref{fig:LEdust} this is likely an incorrect assumption. In other words, the actual color of LE1 might be bluer and  more similar to the segment located in the N direction than what is inferred from integrating the flux over the entire ring. The color from this segment is over-plotted in  the right-hand panel of Fig.~\ref{fig:echo_flux}. Adopting these colors rather than those from the integrated flux of the ring, we infer an average $\tau^V_{eff}$ value of $\sim 0.6\pm0.4$. This is aligned with the range of values adopted by \citet{patat2005}  in their  single scattering plus attenuation model.

We now leverage the results gleaned from the analysis to construct the simple schematic of the 2-D distribution of the dust sheets generating the LEs of SN~2016adj in Fig.~\ref{fig:LEschematic}. The parabolas in the diagram correspond to the iso-delay surfaces of the epochs of HST observations. The appearance of the LE components combined with our inferred $z_{SM}$ values suggests that the dust within the scattering surfaces is distributed non-homogeneously with significant `holes' within the sheets. 

Within this scheme, LE1 is formed by a dense, clumpy layer of CSM located along the LoS extending over an area perpendicular to the iso-delay surfaces of at least $\sim$28$\pm$10 pc. LE3 and LE4 are formed by distinct dust sheets characterized by rather extensive holes  relative to our LoS, and  first encountered by the scattering ellipsoid between +578~d and +1991~d. 
Turning to LE2, although this component exhibits the smallest $z_{SM}$ value it does not appear until late times, which could also be due to a rather extensive hole within the sheet along the LoS. 

\section{Conclusion}
\label{sec:conclusion}
 
We have presented a stunning set of HST images documenting the emergence and evolution of multiple LEs associated with SN~2016adj in Cen~A, along with measurements of the brightness, color, and angular size on the sky of each LE component. Following the single scattering formalism, a 2-D schematic is constructed consisting of four discrete patchy dust sheets producing the echo components. Although beyond the scope of this Letter, one could apply a Heyney–Greenstein phase function analysis to construct an exquisite 3-D mapping of the dust structures along the LoS, and in doing so place constraints on the underlying dust properties \citep[e.g.,][]{yang2017}. Finally, given the complex dust lane of Cen~A, further HST images could not only document the evolution of the late LE components, but also possibly reveal the emergence of additional LEs.

\begin{acknowledgments}
We are grateful to the reviewer for a thorough and timely review which pointed out a handful of errors.
This research is funded by grants from the Independent Research Fund Denmark (8021-00170B) and the Villum FONDEN (28021). MF is supported by a Royal Society - Science Foundation Ireland University Research Fellowship. L.G. acknowledges financial support from the MCIU AEI 10.13039/501100011033 PID2020-115253GA-I00 HOSTFLOWS project, the 2019 Ram\'on y Cajal program RYC2019-027683-I, the PIE project 20215AT016, and the program CEX2020-001058-M. Data presented in the paper were made with the NASA/ESA \textit{Hubble Space Telescope} under program ID: 11360 (PI O'Connell), 14115 (PI Van Dyk), 14487 (PI Sugerman), 14700 (PI Sugerman), and 16179 (PI Filippenko). Images were retrieved from the archive at the Space Telescope Science Institute. STScI is operated by the Association of Universities for Research in Astronomy, Inc. under NASA contract NAS 5-26555. Support for SSL via programs 14487 and 14700 was provided by NASA through a grant from the Space Telescope Science Institute, which is operated by the Association of Universities for Research in Astronomy, Inc., under NASA contract NAS 5-03127.

\end{acknowledgments}

\bibliography{bibtex.bib}

\begin{deluxetable*}{ccccccc}[t] 
\tablewidth{0pt}
\tabletypesize{\scriptsize}
\tablecaption{Log of the HST+WFC3 observations of SN~2016adj.\label{tab:hst_log}}
\tablehead{
\colhead{Epoch} &
\colhead{Image ID} &
\colhead{Proposal nr} &
\colhead{UT date}&
\colhead{Phase$^{\rm (a)}$}&
\colhead{Total integration}&
\colhead{Filter}\\
\colhead{} &
\colhead{} &
\colhead{} &
\colhead{}&
\colhead{(d)}&
\colhead{(s)}&
\colhead{}}
\startdata
0$^{\rm (b)}$& ib6wrap1q & 11360 & 2010-07-17 11:33:33 & $-$2039 & 1605 & F438W\\
             & ib6wrciyq & 11360 & 2010-07-06 13:40:17 & $-$2049 & 1250 & F547M\\
             & ib6wrcjyq & 11360 & 2010-07-06 18:15:43 & $-$2049 & 1240 & F814W\\
             &           &       &                     &           &      &      \\
1            & icvy01010 & 14115 & 2016-02-22 01:26:29 & 7      & 120  & F438W\\
             & icvy01020 & 14115 & 2016-02-22 01:47:16 & 7      & 120  & F438W\\
             & icvy01030 & 14115 & 2016-02-22 01:49:43 & 7      & 40   & F814W\\
             & icvy01040 & 14115 & 2016-02-22 01:51:30 & 7      & 40  & F814W\\
             & icvy01050 & 14115 & 2016-02-22 01:53:17 & 7      & 40  & F814W\\
             & icvy01060 & 14115 & 2016-02-22 01:55:04 & 7      & 40  & F814W\\
             &           &       &                     &           &     &      \\
2            & id3q01010  & 14487 & 2016-03-19 23:15:45 & 34   & 1400 & F438W\\
             & id3q01020  & 14487 & 2016-03-19 23:42:47 & 34   & 100  & F547M\\
             & id3q01030 & 14487 & 2016-03-20 00:38:10 & 34    & 1200 & F547M\\
             & id3q01040  & 14487 & 2016-03-20 01:01:53 & 34   & 1200 & F814W\\
             & id3q01050  & 14487 & 2016-03-20 02:08:28 & 34   & 120  & F814W\\
             & id3q01060  & 14487 & 2016-03-20 02:22:51 & 34   & 300  & F547M\\ 
             &            &       &                     &         &        &    \\
3            & id3q02010  & 14487 & 2016-04-29 15:11:42 & 75    & 1200 & F547M\\
             & id3q02020  & 14487 & 2016-04-29 15:35:28 & 75    & 1388 & F438W\\
             &            &       &                     &         &      &      \\
4            & id3q03010  & 14487 & 2016-06-25 23:11:58 & 132    & 1200 & F547M\\
             & id3q03020  & 14487 & 2016-06-26 00:17:43 & 132    & 1388 & F438W\\
             &             &       &                    &          &      &      \\
5            & id6h04010  & 14700 & 2016-12-12 18:30:25 & 302    & 1600 & F547M\\
             & id6h04020  & 14700 & 2016-12-12 19:00:48 & 302    & 1388 & F814W\\
             & id6h04030  & 14700 & 2016-12-12 20:09:57 & 302    & 2500 & F438W\\
             &            &       &                     &          &      &      \\
6            & id6h05010 & 14700 & 2017-04-01 09:38:47 & 412    & 1600 & F547M\\
             & id6h05020 & 14700 & 2017-04-01 10:52:12 & 412    & 1388 & F814W\\
             & id6h05030 & 14700 & 2017-04-01 11:19:06 & 412    & 2500 & F438W\\
             &           &       &                     &          &      &      \\
7            & id6h06010 &  14700 & 2017-09-15 02:01:24 &  578    & 1600 & F547M\\
             & id6h06020 &  14700 & 2017-09-15 02:31:47 &  578    & 1388 & F814W\\
             & id6h06030 &  14700 & 2017-09-15 03:42:13 &  578    & 2500 & F438W\\
             &           &        &                     &           &      &      \\
8            & ieb310010 & 16179  & 2021-07-28 13:14:53 &  1991   & 780 & F438W\\  
             & ieb310020 & 16179  & 2021-07-28 13:24:06 &  1991   & 720 & F555W\\  
\enddata
\tablenotetext{a}{Days relative to the epoch of $B$-band maximum estimated  to have occurred on JD-$2,457,433.47\pm2.0$. (see Stritzinger et al., in prep).}
\tablenotetext{b}{Epoch 0 corresponds to pre-SN imaging used as host-galaxy templates to subtract host contamination from each science image.}
\end{deluxetable*}

\begin{appendix}

\section{Data analysis supporting material}
\label{sec:dataanalysis}

\subsection{Recovering LE1 on $+$34~d}
\label{sec:recoveringLE}

Fig.~\ref{fig:psfsub} displays the $+34$~d $F438W$-band HST image of SN~2016adj prior to PSF subtraction (left panel) and after PSF subtraction (right panel). The PSF subtracted image reveal LE1, which is the earliest detected LE documented in conjunction with a SN. Aperture photometry of the LE indicates a $\approx$12\% contribution to the total SN$+$ring flux in the F438W filter as measured from the un-subtracted image, while the LE recovered in the coeval $F547M$ image contributes $\approx 2\%$.

\subsection{Procedures to measure the angular radius $\alpha$}
\label{sec:measuringangle}

Measurements of $\alpha$ for each LE component were computed adopting the following procedure. First, the PSF of the SN was removed from the $+$34~d and $+$75~d images to minimize SN contamination, while this was not necessary in the later epoch images. Next, two circles were drawn containing each LE on the host-subtracted images. Then at each PA, the radius corresponding to the maximum intensity within the area delimited by the two circles encapsulating each LE component was determined and the median of these values serves as the best estimate of $\alpha$. The associated uncertainty on $\alpha$ (and $\rho_{ring}$) corresponds to the median of the differences between the best radius fit and the maximum flux of the ring at each position angle. This error is propagated along with the error on the light-curve phase to estimate the uncertainties associated with estimates on  $z_{SM}$.

\subsection{On the orientation of LE1's dust sheet}
\label{sec:LEorientation}

Following \citet{tylenda2004}, a coincident position of the SN and the center of a LE ring suggests the dust slab producing the ring emission is oriented perpendicular to our LoS. To verify whether or not the sheet of dust producing the LE ring associated with LE1 is indeed not inclined to our LoS we performed the following. First, the brightest positions of LE1 in images obtained on $+$302~d and $+$412~d were fit with three different functions. This included a ring with the center fixed at the SN position, a ring with the center as a fit-parameter, and an ellipse with the the center as a free parameter. The functions produce good fits to the brightest regions of the LE with similar reduced chi-squared estimates. The best-fit ring and ellipse functions were both determined to have a center position coincident with that of the position of SN~2016adj to within 0.6$\pm$0.1 and 0.6$\pm$0.3 pixels, respectively. This exercise suggests that the sheet of dust producing LE1 is oriented perpendicular to our LoS.

\subsection{Surface brightness of LE1}

The top row of Fig.~\ref{fig:fluxvsPA} displays the LE flux vs. PA of LE1 as measured from HST images obtained between  $+$75~d to $+$1991~d, while the bottom row plots the corresponding $F438W$ and $F547M$ flux ratios. In cases when the diffraction spike from the bright star to the NW contaminated the LE emission, the aperture flux was extracted as a function of PA and a correction was determined for the LE flux after removing the contamination by linear interpolation over the PA. 

\subsection{Estimating $\Delta z$ }
\label{sec:deltaz}

Following \citet[][their Eq.~11]{Sugerman2003}, the thickness ($\Delta z_{SM}$) of a dust sheet producing a LE  can be estimated in terms of $\rho$, $\Delta \rho$, $t$, and the light-curve width parameter $\tau$. Estimates of $\Delta \rho$ were inferred from the full-width-at-half-maximum (FWHM) measurements of the rings in the $F438W$-band images after integrating the ring flux at different radii over 360 degrees. Depending on the epoch and on the LE, we obtained FWHM values between $\sim$3-4 pixels. Assuming $\tau = 30$~d, $\Delta z_{SM}$ values associated with the dust sheet producing LE1 from   $+$132~d and $+$578~d range between $\Delta z_{SM}=68\pm2$~pc  $\Delta z_{SM}=117\pm2$~pc. Here the associated errors accounts for the uncertainty on the measured FWHM values of each ring. The LEs observed at the last epoch show FWHM of the rings between 2 and 4 pixels, corresponds to values of $\Delta z_{SM} = 16 \pm 5$ pc for LE2, $z_{SM} = 32\pm11$ pc for LE3, and $z_{SM} =38 \pm 13$ pc for LE4. All  $z_{SM}$ and $\Delta z_{SM}$ estimates are provided in Table~\ref{tab:radiuslog}.

\subsection{RGB image of LE1 of SN~2016adj}
\label{sec:LEdust}

Composite RGB (red, green blue) images constructed by aligning and stacking HST images obtained on $+302$~d, $+412$~d, and $+578$~d are shown in the top row of Fig.~\ref{fig:LEdust}. The bottom row displays the same images with superposed the intensity of contours corresponding to the $F438W$-band flux, with regions lacking contours tracing the complex dust lanes extending across the field.

A prevalent dust lane running in the  E-NE to W-SW direction and across  LE1, reveals significant attenuation of both background stars and the LE1 as it radially expands on the sky. Clearly there is a significant amount of dust at very large distance ($>$ 1,000 pc) in front of the position of SN~2016adj, and has a large thickness along the z-axis. This host dust could be a single structure or multiple layers, and surely contributes to the large host visual extinction inferred from the analysis of the SN colors (Stritzinger M. et al., in prep). This extended dust structure could still allow for a relatively low density of dust in the close-in sheet producing LE1, allowing that dust to remain in the single scattering regime. The photons we are seeing as LE1 all traveled in a straight line from the SN to the dust sheet producing LE1, and only suffered one scattering there to be directed into our LoS to Earth. The distant foreground dust is dimming these LE photons, but if it is far enough in the foreground and extended over a long depth along the LoS the LE it would produce from photons that traveled directly from the SN would be faint, at large radii, and diffusely smeared out and possibly --too faint to be detected.

\renewcommand\thefigure{\thesection.\arabic{figure}}  
\setcounter{figure}{0}

\begin{figure*}[htb]
\centering
\includegraphics[width=16cm]{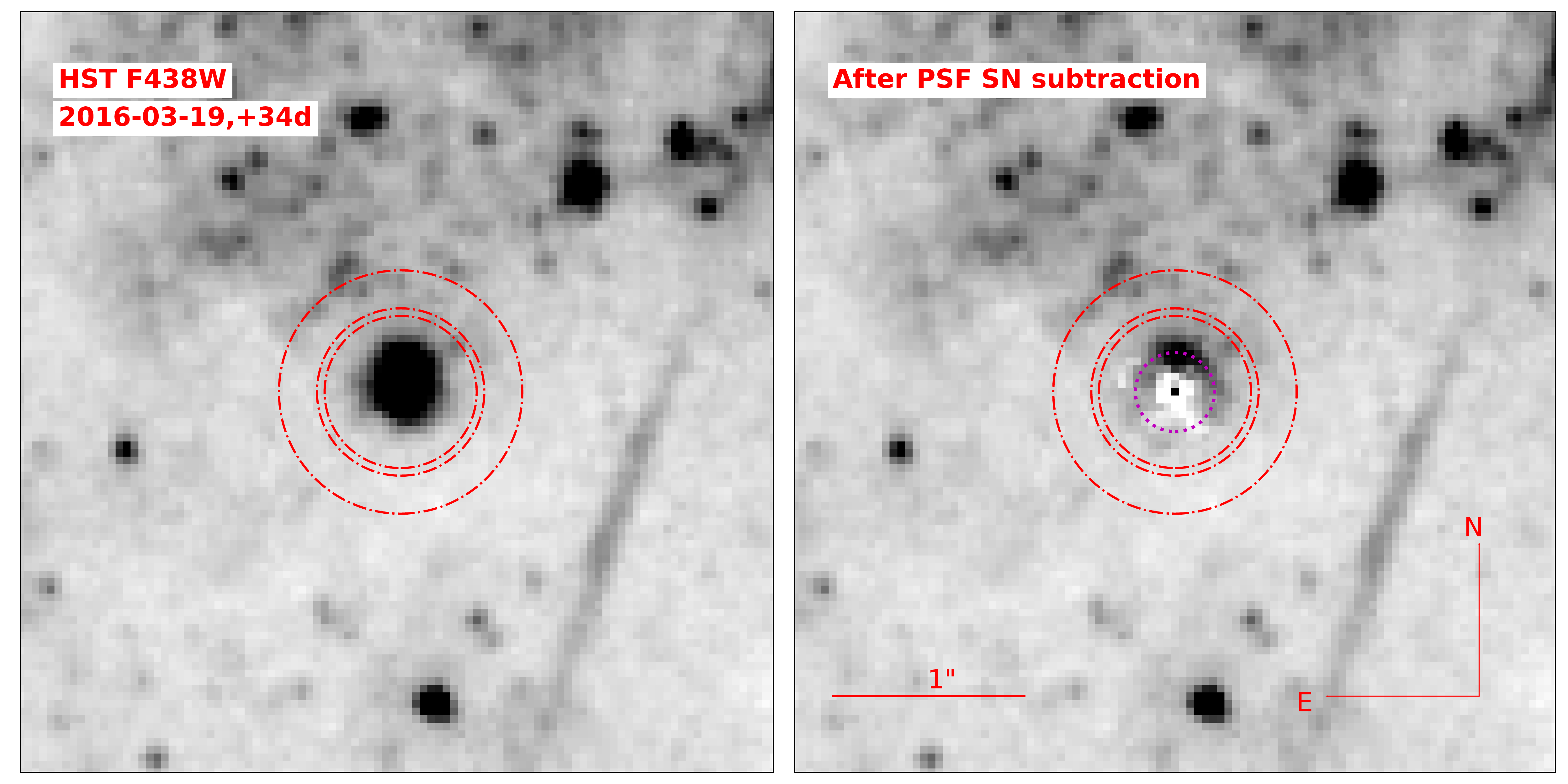}
\caption{\textit{Left:} HST $F438W$-band image of SN~2016adj obtained on $+$34~d. \textit{Right:} Residual image after subtraction of a PSF model constructed from two-dozen stars. The emerging LE is indicated with a dotted magenta circle. Dashed-dotted red circles correspond to photometry and sky-background apertures. Scale and orientation of the images are reported in the right-hand panel.}
\label{fig:psfsub}
\end{figure*}

\begin{figure*}[!htb]
\centering
\includegraphics[width=18cm]{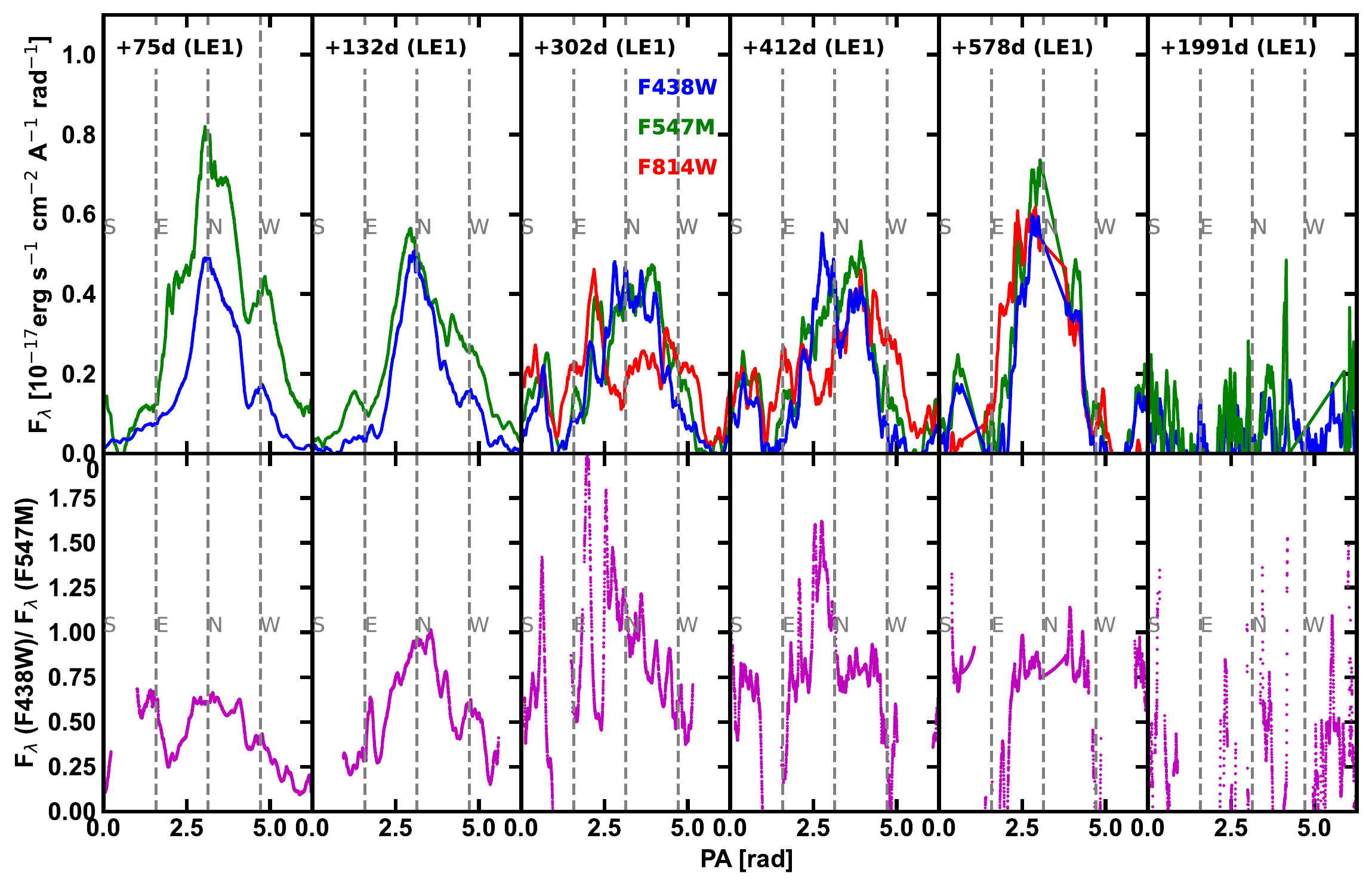}
\caption{\textit{Top:} Fluxes of LE1 over six epochs plotted as a function of PA in two and/or three HST filters. \textit{Bottom:} Ratio of $F438W$- and $F547M$-band flux measurements as a function of PA.}
\label{fig:fluxvsPA}
\end{figure*}

\begin{figure*}[!htb]
\centering
\includegraphics[width=18cm]{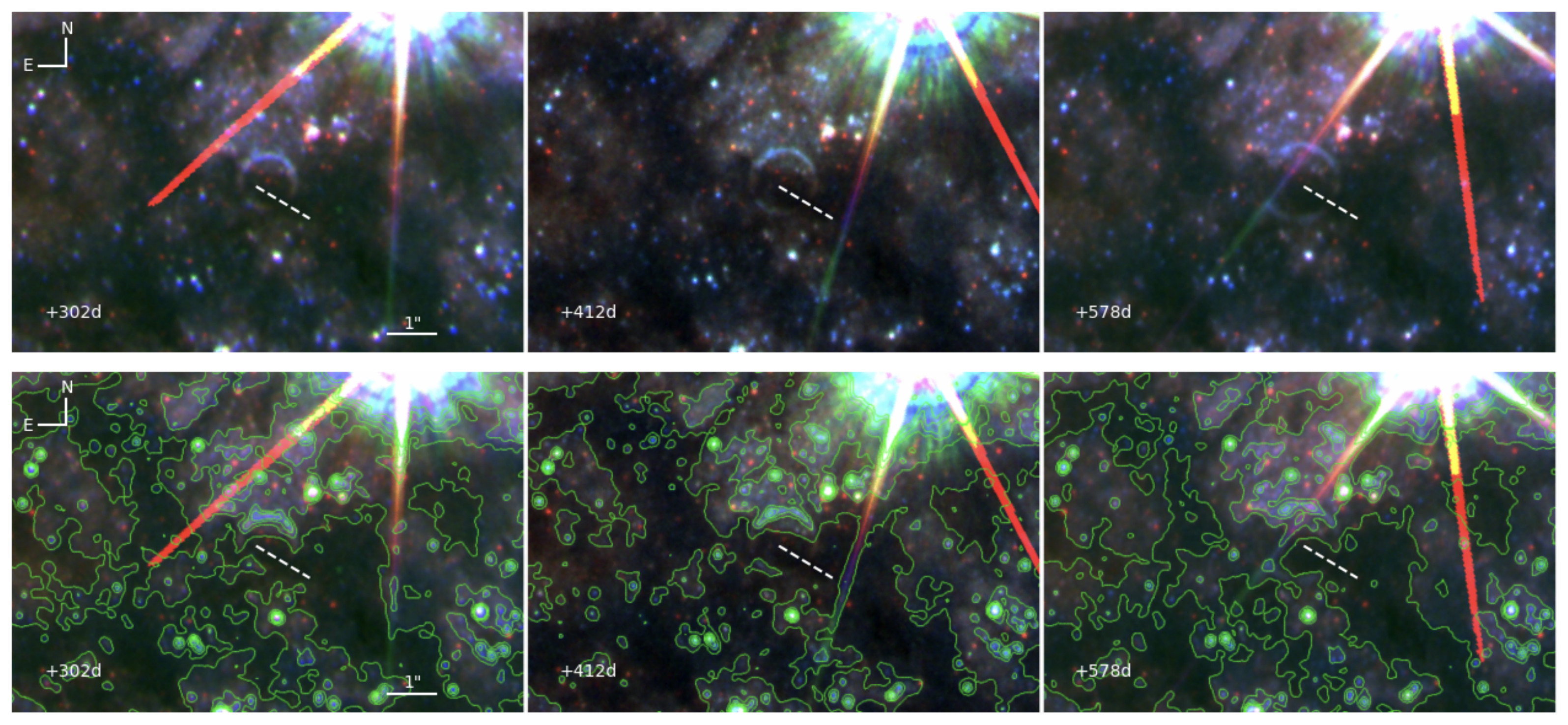}
\caption{\textit{Top:} Composite RGB images centered on LE1 from epochs between $+302$~d and $+578$~d. 
A prominent dust lane indicated by a dashed white line and running E-NE to W-SW in the direction across the ring of LE1 clearly  obscures both background stars and LE1  as it radially expands on the sky.  The apparent color inferred from the total integrated flux is redder than the intrinsic color, which is more accurately traced by the segment of the ring extending from the N direction (see Fig.~\ref{fig:fluxvsPA}). \textit{Bottom:} Same images  as in the top row but with  contour lines superposed corresponding to different  $F438W$-band flux levels. Regions of minimal contours trace the spectacular dust lanes covering the field of SN~2016adj.}
\label{fig:LEdust}
\end{figure*}

\newpage
\setcounter{table}{0}
\renewcommand{\thetable}{A\arabic{table}}
\begin{deluxetable*}{lccllclllccccc}
\rotate
\tabletypesize{\scriptsize}
\tablewidth{0pt}
\tablecaption{Light-echo properties\label{tab:radiuslog}}
\tablehead{
\colhead{Epoch} &
\colhead{phase$^{\rm (a)}$} &
\colhead{Angular Radius}&
\colhead{$\rho_{ring}$}&
\colhead{$z_{SM}$} &
\colhead{$\Delta z_{SM}$} &
\colhead{F$_{F438W}$} &
\colhead{m$_{F438W}$} &
\colhead{F$_{F547M}$} &
\colhead{m$_{F547M}$} &
\colhead{F$_{F814W}$} &
\colhead{m$_{F814W}$} &
\colhead{$\tau^V_{eff}$ring$^{\rm (b)}$}&
\colhead{$\tau^V_{eff}$N$^{\rm (c)}$}\\
\colhead{} &
\colhead{(d)} &
\colhead{$\alpha$(\arcsec)}&
\colhead{(pc)} &
\colhead{(pc)} &
\colhead{(pc)} &
\colhead{(**)} &
\colhead{(mag)} &
\colhead{(**)} &
\colhead{(mag)} &
\colhead{(**)} &
\colhead{(mag)}& 
\colhead{} &
\colhead{} }
\startdata
2 (LE1)       &  +34          & 0.19$\pm$0.03  &  3.18$\pm$0.53  & 176.68$\pm$60.44 &  \ldots    & 0.674$\pm$0.135 & 22.34$\pm$0.22 & 0.770$\pm$0.154  & 21.70$\pm$0.22 & \ldots   & \ldots  &  $1.0\pm$1.0 & \ldots \\
3 (LE1)       &  +75          & 0.27$\pm$0.01  &  4.44$\pm$0.18  & 157.21$\pm$13.74 & \ldots & 0.993$\pm$0.013 & 21.92$\pm$0.01 & 2.069$\pm$0.013 & 20.63$\pm$0.01 & \ldots   & \ldots  &   $2.8\pm$0.2 & 2.0$\pm$0.4 \\
4 (LE1)       & +132          & 0.38$\pm$0.02  &  6.27$\pm$0.40  & 177.41$\pm$22.79 & 117$\pm$2 & 0.884$\pm$0.013 & 22.05$\pm$0.02 & 1.377$\pm$0.013 & 21.07$\pm$0.01 & \ldots   & \ldots  & $1.9\pm$0.2 &  0.6$\pm$0.3 \\
5 (LE1)       & +302          & 0.56$\pm$0.04  &  9.22$\pm$0.60  & 167.80$\pm$21.79 & 90$\pm$2  & 1.062$\pm$0.021 & 21.85$\pm$0.02 & 1.253$\pm$0.019 & 21.17$\pm$0.02 & 1.20$\pm$0.02 & 20.38$\pm$0.02  & $1.1\pm$0.2 & 0.1$\pm$0.2 \\
6 (LE1)       & +412          & 0.66$\pm$0.03  & 10.94$\pm$0.53  & 172.97$\pm$16.84 & 68$\pm$2 & 1.019$\pm$0.021 & 21.89$\pm$0.02 & 1.252$\pm$0.025 & 21.18$\pm$0.02 & 1.30$\pm$0.03 & 20.29$\pm$0.02 & $1.2\pm$0.2 &  0.3$\pm$0.2\\
7 (LE1)       & +578          & 0.79$\pm$0.04  & 13.10$\pm$0.60  &  176.56$\pm$16.14 & 76$\pm$2  & 1.117$\pm$0.027 & 21.79$\pm$0.03 & 1.453$\pm$0.054 & 21.01$\pm$0.04 & 1.29$\pm$0.14 & 20.31$\pm$0.11 & $1.3\pm$0.3 & 1.4$\pm$0.4\\ 
8 (LE1)       & +1991$^{*}$   & 1.47$\pm$0.08  & 24.31$\pm$1.31  & 175.98$\pm$19.12 &  28$\pm$10 & 0.221$\pm$0.051 & 23.55$\pm$0.25 & 0.411$\pm$0.027 & 22.39$\pm$0.07 & \ldots   & \ldots  & 2.4$\pm$0.9 & \ldots \\
\hline                                             
8 (LE2)       & +1991$^{*}$   & 0.87$\pm$0.08  & 14.35$\pm$1.33  & 60.76$\pm$11.40 & 16$\pm$5 & 0.289$\pm$0.041 & 23.26$\pm$0.16 & 0.371$\pm$0.049 & 22.50$\pm$0.14 & \ldots   & \ldots   & 1.3$\pm$0.7 & \ldots \\
\hline                                                         
8 (LE3)       & +1991$^{*}$   & 1.70$\pm$0.12  & 28.25$\pm$1.97  & 237.98$\pm$33.32 & 32$\pm$11 & 0.242$\pm$0.059 & 23.45$\pm$0.27 & 0.421$\pm$0.032 & 22.36$\pm$0.08 & \ldots   & \ldots   & $2.2\pm$0.9 & \ldots \\
\hline                         
8 (LE4)       & +1991$^{*}$   & 1.98$\pm$0.12  & 32.85$\pm$1.97  & 322.06$\pm$38.75 & 38$\pm$13 & 0.188$\pm$0.068 & 23.72$\pm$0.41 & 0.301$\pm$0.037 & 22.72$\pm$0.13 & \ldots   & \ldots   & $2.0\pm$1.4 & \ldots \\   
\enddata
\tablecomments{The uncertainties in the flux measurements of the first two epochs include a conservative $20\%$ uncertainty given the contaminating residuals created by our PSF SN subtraction. 
The $F438W$, $F547M$, and $F814W$ filters are similar to ground-based $BVi$ filters. The magnitudes are reported in the AB system.}
\tablenotetext{$a$}{Days past the epoch of $B$-band maximum.}
\tablenotetext{$b$}{Values computed adopting colors computed from full LE ring. The quoted uncertainties account for 0.05 mag error in the adopted intrinsic SN peak color, and the observational errors of the apparent color of the ring.}
\tablenotetext{$c$}{Values computed adopting colors computed from the Northern bright segment of   LE1.}
\tablenotetext{$**$}{10$^{-17}$ erg s$^{-1}$ cm$^{-2}$ \textrm{\AA}$^{-1}$.}
\tablenotetext{$*$}{For the last epoch measurements are reported for LE1, LE2, LE3, and LE4. At this epoch the filter $F547M$ is actually $F555W$.}

\end{deluxetable*}

\end{appendix}
\end{document}